\documentclass[aps,preprint,preprintnumbers,amsmath,amssymb,nofootinbib]{revtex4}
\usepackage{amssymb}
\usepackage{amsmath}
\usepackage{bm}
\usepackage{graphicx}
\usepackage{epsfig}
\begin{document}
\title{Pre-big bang collapsing universe from modern Kaluza-Klein theory of gravity.}
\author{$^{1,2}$ Mauricio Bellini \footnote{ E-mail address:
mbellini@mdp.edu.ar, mbellini@conicet.gov.ar} } \vskip .2cm
\address{$^1$ Departamento de F\'isica, Facultad de Ciencias Exactas y
Naturales, Universidad Nacional de Mar del Plata, Funes 3350,
C.P. 7600, Mar del Plata, Argentina.\\  \\
$^2$ Instituto de Investigaciones F\'{\i}sicas de Mar del Plata (IFIMAR), \\
Consejo acional de Investigaciones Cient\'ificas y T\'ecnicas
(CONICET), Argentina.}

\begin{abstract}
We study the collapse of the universe described by a scalar field
spherically symmetric collapse of a system described by a massless
scalar field  from a 5D Riemann-flat canonical metric, on which we
make a dynamical foliation on the extra space-like dimension. The
asymptotic universe (absent of singularities) results to be finite
in size and energy density, with an vacuum dominated equation of
state. The important result here obtained is that the asymptotic
back-reaction effects are given by a negative constant:
$\left.\frac{1}{2}
\left[\frac{1}{1+\dot\psi^2}+\frac{1}{\dot\psi^2}\right]
\left<\left(\dot{\delta\bar{\varphi}}\right)^2 \right>+{1\over
2a^2} \left< \left(\vec{\nabla}\delta\varphi\right)^2
\right>\right|_{t\rightarrow\infty} = {-8 \Lambda_0 \over 3 \pi
G}$.
\end{abstract}
\maketitle

\section{Introduction, basic equations and motivation}

The cosmological models based on classical General Relativity (GR)
predict that the Big Bang singularity is the true beginning of the
universe (boundary of the spacetime). It has been long expected
that the existence of singularity in the classical general
relativity which has been shown to be quite generic thanks to the
singularity theorems, will be removed when classical framework of
gravity is extended to a quantum framework of gravity. When one
describes the early universe, GR is applied beyond its domain of
validity. The quantum effects which dominate in this epoch are
expected to resolve the singularity. In particular, the existence
of the cosmological singularity in the framework of Loop Quantum
Gravity (LQG)\cite{4} has been subject of study in the last years.
During the inflationary big bang expansion, the universe suffered
an exponential accelerated expansion driven by a scalar (inflaton)
field with an equation of state close to a vacuum dominated one.
The most conservative assumption is that the energy density $\rho
=P/\omega $ is due to a cosmological parameter which is constant
and the equation of state is given by a constant $\omega =-1$,
describing a vacuum dominated universe with pressure $P$ and
energy density $\rho$. On the other hand, exists a kind of exotic
fluids that may be framed in theories with matter fields that
violate the weak energy condition\cite{u}, such that $\omega <-1$.
These models were called phantom cosmologies, and their study
represents a currently active area of research in theoretical
cosmology\cite{uu}.

On the other hand, the spherically symmetric collapse of a
massless scalar field has been of much interest towards
understanding the dynamical evolutions in general relativity.
Both, analytical\cite{[1]} and numerical\cite{[2]} investigations,
have been undertaken by various authors to gain more insight into
the formation of black holes. One remarkable finding of these
numerical investigations is the demonstration of criticality in
gravitational collapse. Specifically, it was found that for a
range of values of the parameter characterizing the solution,
black hole forms and there was a critical value of the parameter
beyond which the solutions are such that the scalar field
disperses without forming any black hole. However, this result has
been obtained mainly through numerical studies and a proper
theoretical understanding of this phenomenon is still lacking (see
e.g. \cite{[3]} and the references therein).

In this letter we study the cosmological collapse of a pre-big
bang universe driven by a massless scalar field in the framework
of the 5D Modern Kaluza-Klein theory of gravity\footnote{This
topic has been studied earlier by Liu and Wesson\cite{5}.}, also
known as the Space-Time-Matter theory of gravity\cite{wo,we}. We
shall use a dynamical foliation, $\psi\equiv \psi(t)$, of the
noncompact space-like extra dimension $\psi$. This topic was
explored previously by Ponce de Le\'on\cite{pdl}. He showed that
the FRW line element can be "reinvented" on a dynamical
four-dimensional hypersurface, which is not orthogonal to the
extra dimension, without any internal contradiction. The effective
4D hypersurface is selected by the requirement of continuity of
the metric and depends explicitly on the evolution of the extra
dimension. More recently, was demonstrated that phantom
scenarios\cite{pc}, warm inflation\cite{ruso} and super
exponential inflationary scenarios\cite{ul} can be obtained
through this mechanism from a 5D Riemann-flat.

In order to study the dynamics of a scalar field $\varphi$ on a 5D
vacuum, we consider the canonical metric
\begin{equation}\label{m}
dS^2 = g_{\mu\nu} (y^{\sigma},\psi) d y^{\mu} dy^{\nu} - d\psi^2.
\end{equation}
Here the 5D coordinates are orthogonal: $y \equiv
\{y^a\}$\footnote{Greek letters run from $0$ to $3$, and latin
letters run from $0$ to $4$.}. The geodesic equations for an
relativistic observer are
\begin{equation}\label{geo}
\frac{dU^a}{dS} + \Gamma^a_{bc}\,\,U^b U^c =0,
\end{equation}
where $U^a = {dy^a\over dS}$ are the velocities and
$\Gamma^a_{bc}$ are the connections of (\ref{m}). Now we consider
a parametrization $\psi(x^{\alpha})$, where $x\equiv
\{x^{\alpha}\}$ are an orthogonal system of coordinates, such that
the effective line element (\ref{m}), now can be written as
\begin{equation}
dS^2 = h_{\alpha\beta}\,dx^{\alpha} dx^{\beta}.
\end{equation}
It is very important to notice that $S$ will be an invariant, so
that derivatives with respect to $S$ will be the same on 5D or 4D.
In other words, in this paper we shall consider that spacetime
lengths that remain unaltered when we move on an effective 4D
spacetime.

We are interested to study how is the effective 4D dynamics of
$\varphi$, obtained from a dynamical foliation of a 5D Ricci-flat
canonical metric. We consider a classical massless scalar field
$\varphi(y^a)$ on the metric (\ref{m}). In order to make a
complete description for the dynamics of the scalar field, we
shall consider its energy momentum tensor. In order to describe a
true 5D physical vacuum we shall consider that the field is
massless and there is absence of interaction on the 5D Ricci-flat
manifold, so that
\begin{equation}
T^a_{\,\,b} = \Pi^a \Pi_b - g^a_{\,\,b} \,{\cal
L}\left[\varphi,\varphi_{,c}\right],
\end{equation}
where ${\cal L}\left[\varphi,\varphi_{,c}\right] = {1\over 2}
\varphi^a\varphi_a$ is the lagrangian density for a free and
massless scalar field on (\ref{m}) and the canonical momentum is
$\Pi^a = {\partial {\cal L}\over
\partial \varphi_{,a}}$. Notice that we are not considering interactions on the 5D vacuum,
because it is related to a physical vacuum in the sense that the
Einstein tensor is zero: $G^a_b =0$. The effective 4D energy
momentum tensor will be
\begin{equation}
\bar{T}_{\alpha\beta} = \left.e^a_{\,\,\alpha} \, e^b_{\,\,\beta}
\,T_{ab}\right|_{\psi(x^{\alpha})}.
\end{equation}
In other words, using the fact that ${\cal L}$ is an invariant it
is easy to demonstrate that
\begin{equation}
\bar{T}^{\alpha}_{\,\,\beta} = \bar{\Pi}^{\alpha}\,
\bar{\Pi}_{\beta} - h^{\alpha}_{\,\,\beta} {\cal L},
\end{equation}
where ${\cal L}$ is an invariant of the theory
\begin{equation}\label{ao}
{\cal L} = \frac{1}{2} \varphi^{,a} \varphi_{,a} = \frac{1}{2}
\left(e^a_{\alpha} \bar{\varphi}^{,\alpha}\right)
\left(\bar{e}^{\beta}_a \bar{\varphi}_{,\beta}\right) .
\end{equation}

\section{Collapsing universe: basic equations}

We consider the 5D Riemann-flat metric\cite{eu}
\begin{equation}\label{m1}
dS^2 = \psi^2 \left[ \frac{\Lambda(t)}{3} dt^2 - e^{-2\int
\left(\frac{\Lambda(t)}{3}\right)^{1/2}\, dt} \, dr^2 \right] -
d\psi^2,
\end{equation}
which describes 5D extended universe with variable cosmological
function $\Lambda(t)>0$, which is contracting with the time. The
noncompact extra coordinate $\psi$ has spatial unities (we shall
consider unities $c=\hbar=1$). Furthermore, $dr^2 = dx^2 + dy^2 +
dz^2$ is the 3D Euclidean metric, t is the cosmic time and $\psi$
is the space-like noncompact extra dimension. Since the metric (1)
is Riemann-flat (and therefore Ricci-flat), hence it is suitable
to describe a 5D vacuum vacuum ($G_{ab}=0$) in the framework of
Space-Time-Matter (STM) theory of gravity\cite{6}. With this aim
we shall consider the 5D action
\begin{equation}\label{act}
^{(5)}I = {\Large\int} d^4 x \  d\psi \sqrt{\left|\frac{^{(5)}
 g}{^{(5)} g_0}\right|} \left(
\frac{^{(5)} R}{16\pi G}+ \frac{1}{2}g^{ab} \varphi_{,a}
\varphi_{,b} \right),
\end{equation}
where $^{(5)}g$ is the determinant of the covariant metric tensor
$g_{AB}$:
\begin{equation}\label{ricci5}
^{(5)}g= \psi^8 \,\left(\frac{\Lambda}{3}\right) \, e^{-6\int
\sqrt{\frac{\Lambda}{3}} dt},
\end{equation}
and $^{(5)} g_0=\psi^8_0 \left({\Lambda_0\over3}\right)$ is a
constant to make dimensionless the expression $\left|^{(5)} g
/^{(5)} g_0\right|$.

In order to describe a 5D vacuum on (\ref{m1}), we shall consider
$\varphi$ as a massless test classical scalar field, which is
minimally coupled to gravity. For this reason, the
$\varphi$-contribution to the Lagrangian will be considered as
purely kinetic and free of any interaction on (\ref{m1}). From the
mathematical point of view, the second term in the action
(\ref{act}) is constructed by using monogenic fields $\varphi$,
which have null D\'{}Alambertian on the 5D Riemann-flat metric
(\ref{m1})\cite{almeida}.

Now we consider a dynamical foliation: $\psi\equiv \psi(t)$, on
the metric (\ref{m1}). If we require that $t$ to be a cosmic time,
we need
\begin{equation}\label{time}
\psi^2 \frac{\Lambda(t)}{3} - \dot\psi^2 =1,
\end{equation}
and hence
\begin{equation}
\Lambda(t) = 3 \left[ \frac{1+\dot\psi^2}{\psi^2}\right].
\end{equation}
In this paper we shall study the particular case where
$\Lambda(t)$ is a constant of time: $\Lambda(t)=\Lambda_0$. In
this particular case the solutions are
\begin{equation}\label{...}
\psi_{(D,I)}(t)= \frac{1}{6} \left[ \frac{9 + e^{\mp
2\sqrt{{\Lambda_0\over 3}} t}}{\sqrt{{\Lambda_0\over 3}} \, e^{\mp
\sqrt{{\Lambda_0\over 3}}t}}\right],
\end{equation}
where the solution $\psi_{(D)}(t)$ decreases with time and
$\psi_{(I)}(t)$ increases. The expanding version of this solution
was considered in\cite{abc}. In this letter we shall consider the
model generated with $\psi_{(D)}(t)$. To simplify the notation we
shall call it: $\psi_{(D)}(t)\equiv \psi(t)$\footnote{Note that
the large times value of $\psi(t)$ is $\psi_0\equiv
\left.\psi(t)\right|_{t\rightarrow \infty} \rightarrow 3/2$}. In
this case the scale factor of the universe on the 4D hypersurface
described by the effective 4D metric
\begin{equation}\label{4d}
dS^2 = dt^2 - \psi^2(t) \, e^{-2 \sqrt{{\Lambda_0\over 3}} t} \,
dr^2,
\end{equation}
is
\begin{equation}
a(t) = \psi(t) \, e^{-\sqrt{{\Lambda_0 \over 3}} t}.
\end{equation}
Notice that this scale factor tends to a constant as $t$ tends to
infinity
\begin{equation}
a_0\equiv \lim_{t\rightarrow \infty} {a(t)} = \frac{3}{2}
\sqrt{{3\over \Lambda_0}}.
\end{equation}
This is a very interesting behavior because the model describes a
contracting universe which has an asymptotic finite size $a_0$.
This effect is due to the fact in General Relativity the action is
invariant under time reflections. Thus, to any standard
cosmological solution $H(t)$, describing decelerated expansion and
decreasing curvature ($H > 0$, $\dot{H} < 0$), time reversal
associates a "reflected" solution, $H(-t)$, describing a
contracting Universe. In a string cosmology context, this
solutions are called dual\cite{gv}. In this letter we are dealing
with an extra dimensional cosmological model where the extra
dimension is non-compact. However, this duality is preserved and
the interpretation of the results obtained by Gasperini and
Veneziano in \cite{gv} are preserved.

The effective Hubble parameter is given by
\begin{equation}
H(t)= \frac{\dot{a}}{a}= -2 \sqrt{\frac{\Lambda_0}{3}}
\,\frac{e^{-2\sqrt{{\Lambda_0\over
3}}t}}{\left[9+e^{-2\sqrt{{\Lambda_0\over3}}t}\right]} <0,
\end{equation}
which is negative in agreement with that one expects for a
collapsing universe. Notice that the asymptotic Hubble parameter
is $\lim_{t\rightarrow \infty} H(t) \rightarrow 0$.

On the other hand, the relevant components of the Einstein tensor
are (we use cartesian coordinates)
\begin{eqnarray}
G^0_{\,\,0} & = & -\frac{4 \Lambda_0 e^{-4\sqrt{{\Lambda_0\over
3}}t} }{\left(9+e^{-2\sqrt{{\Lambda_0\over 3}}t}\right)^2}, \label{29} \\
G^x_{\,\,x} & = & -\frac{4 \Lambda_0 e^{-2\sqrt{{\Lambda_0\over
3}}t} \left[6+e^{-2\sqrt{{\Lambda_0\over
3}}t}\right]}{\left(9+e^{-2\sqrt{{\Lambda_0\over 3}}t}\right)^2},
\label{30}
\end{eqnarray}
so that, using the fact that the Einstein equations are
respectively $G^{0}_{\,\,0} = -8\pi G\,\rho $ and $G^x_{\,\,x} =
G^y_{\,\,y}= G^z_{\,\,z}= 8\pi G \, P$, we obtain the equation of
state for the universe
\begin{equation}\label{state}
\frac{P}{\rho} = \omega(t) = - e^{2\sqrt{{\Lambda_0\over 3}}t}
\left[ 6 + e^{-2\sqrt{{\Lambda_0\over 3}}t}\right],
\end{equation}
which always remains with negative values $\omega(t) <-1$, and
tends to negative infinity values for large asymptotic times. This
is because the energy density tends to zero very much rapidly than
the pressure. Notice that in a pre-big bang followed with a
post-big bang with time reflexion $t \rightarrow -t$ in the
equation of state (\ref{state}), such that equation would describe
an inflationary post-big bang expansion with
$\left.\omega\right|_{t\rightarrow -\infty} \rightarrow -1$, which
assures a asymptotic spatially flat universe, in agreement with
observations\cite{wmap7}. The effective 4D scalar curvature
\begin{equation}
{\cal{\bar{R}}} = 8 \Lambda_0 \frac{e^{-2\sqrt{\frac{\Lambda_0}{3}
t}}}{\left[ 9 + e^{-2\sqrt{{\Lambda_0\over 3}}}t\right]},
\end{equation}
decreases with the time and has a null asymptotic value
$\left.\cal{\bar{R}}\right|_{t\rightarrow \infty} \rightarrow 0 $.

The expectation values for the energy density and the pressure,
written in terms of the scalar field
$\varphi(t,\vec{r},\psi(t))\equiv \bar{\varphi}(t,\vec{r})$, are
\begin{eqnarray}
\rho = \left<0|\bar{T}^0_{\,\,0} |0\right> & = & \left<
\frac{3}{2\Lambda_0 \psi^2(t)} \dot{\bar{\varphi}}^2 + \frac{1}{2
a^2(t)} \left(\vec{\nabla} \bar{\varphi}\right)^2+ \frac{1}{2}
 \left(\frac{\partial\bar{\varphi}}{\partial \psi}\right)^2 \right>, \label{34} \\
P  = -\left<0| \bar{T}^i_{\,\,j}|0\right> & = & - \delta^i_{\,\,j}
\left<\frac{3}{2\Lambda_0 \psi^2(t)} \dot{\bar{\varphi}}^2 -
\frac{1}{6 a^2(t)} \left(\vec{\nabla} \bar{\varphi}\right)^2 -
\frac{1}{2}
 \left(\frac{\partial\bar{\varphi}}{\partial \psi}\right)^2 \right>. \label{35}
\end{eqnarray}
Here, the notation $\left<0| ... |0\right>$ denotes the quantum
expectation value calculated on a 4D vacuum state. Because we are
considering a spatially isotropic and homogeneous background, we
shall consider an averaging value with respect to a gaussian
distribution on a Euclidean 3D volume.

\section{Background dynamics and back-reaction effects}

We consider the semiclassical expansion for the scalar field
$\bar{\varphi}(\vec{r},t) = \bar{\phi}(t,\psi(t)) +
\delta{\bar{\varphi}}(\vec{r},t)$, such that the expectation value
for the effective 4D scalar field fluctuations becomes null:
$\left<\delta\bar{\varphi}\right> =0$. In general, since the
averaging is considered as gaussian, the  momentums of odd order
in the fluctuations
$\left<\left[\dot{\delta{\bar{\varphi}}}(\vec{r},t)\right]^{(2n+1)}\right>$,
with $n$ integer, will be zero.

Taking into account the fact that
$\left.{\partial\bar{\varphi}\over\partial\psi}\right|_{\psi(t)} =
\dot{\bar{\varphi}}/\dot\psi$, we obtain the effective background
Energy-Momentum components on the 4D effective hypersurface
(\ref{4d})
\begin{eqnarray}
\left<\bar{T}^0_{\,\,0} \right> = \rho & = & \frac{3}{2\Lambda_0
\psi^2(t)} \dot{\bar{\phi}}^2+\frac{3}{2 \Lambda_0 \psi^2(t)}
\left<\left(\dot{\delta\bar{\varphi}}\right)^2 \right> +\frac{1}{2
a^2(t)} \left<\left(\vec{\nabla}
\delta\bar{\varphi}\right)^2\right>  \nonumber \\
&+& \frac{1}{2} \left(\frac{\dot{\bar{\phi}}}{\dot\psi}\right)^2
+\frac{1}{2}
\frac{\left<\left(\dot{\bar{\delta\varphi}}\right)^2\right>}{\dot\psi^2}, \label{41}\\
- \left< \bar{T}^x_{\,\,x}\right> = P & = & \frac{3}{2\Lambda_0
\psi^2(t)} \dot{\bar{\phi}}^2 + \frac{3}{2 \Lambda_0 \psi^2(t)}
\left<\left(\dot{\delta\bar{\varphi}}\right)^2 \right> -\frac{1}{6
a^2(t)} \left<\left(\vec{\nabla}
\delta\bar{\varphi}\right)^2\right>  \nonumber \\
&-& \frac{1}{2}
\left(\frac{\dot{\bar{\phi}}}{\dot\psi}\right)^2-\frac{1}{2}
\frac{\left<\left(\dot{\bar{\delta\varphi}}\right)^2\right>}{\dot\psi^2}
. \label{42}
\end{eqnarray}
Using the Einstein equations with the expression (\ref{time}), we
obtain the important result that describes the temporal evolution
for the background back-reaction effects
\begin{eqnarray}
\frac{1}{2}
\left[\frac{1}{1+\dot\psi^2}+\frac{1}{\dot\psi^2}\right]
\left<\left(\dot{\delta\bar{\varphi}}\right)^2 \right> &+
&\frac{1}{2 a^2}
\left<\left(\vec\nabla{\bar{\delta\varphi}}\right)^2\right> =
\frac{36 \Lambda_0}{\pi G} \left[9+e^{-2\sqrt{{\Lambda_0\over
3}}t}\right]^2  \nonumber \\
& \times & \left\{ e^{-4\sqrt{{\Lambda_0\over 3}}t}
-\frac{\left(81 e^{2\sqrt{{\Lambda_0\over
3}}t}+e^{-2\sqrt{{\Lambda_0\over
3}}t}\right)\left(9+e^{-2\sqrt{{\Lambda_0\over 3}}t}\right)}{2
\left(81 e^{2\sqrt{{\Lambda_0\over 3}}t} -126 +
e^{-2\sqrt{{\Lambda_0\over 3}}t}\right)} \right\}. \label{imp}
\end{eqnarray}

Finally, we obtain the temporal evolution for the background field
\begin{equation}
\dot{\bar{\phi}}^2 = \frac{\Lambda_0 \, }{12\pi G} \frac{\left(
e^{-\sqrt{{\Lambda_0\over 3}}t} - 9 e^{\sqrt{{\Lambda_0\over
3}}t}\right)^2 \left(9 + e^{-2\sqrt{{\Lambda_0\over
3}}t}\right)}{\left(81 e^{2\sqrt{{\Lambda_0\over 3}}t} - 126 +
e^{-2\sqrt{{\Lambda_0\over 3}}t}\right)},
\end{equation}
which, for very large times tends to
 $\left.\dot{\bar{\phi}}^2\right|_{t\rightarrow
\infty} \rightarrow {3 \Lambda_0 \over 4 \pi G} $. On the other
hand, using the equation (\ref{imp}), we obtain the large times
rugosity term
\begin{equation}\label{br}
\frac{1}{2}
\left[\frac{1}{1+\dot\psi^2}+\frac{1}{\dot\psi^2}\right]
\left<\left(\dot{\delta\bar{\varphi}}\right)^2
\right>+\left.\frac{1}{2 a^2}
\left<\left(\vec\nabla{\bar{\delta\varphi}}\right)^2\right>\right|_{t\rightarrow
\infty} = -\frac{8 \Lambda_0}{3\pi G},
\end{equation}
which takes a constant negative value and provide us the
contribution of the squared field fluctuations and their gradients
to the cosmological constant.

\section{Final Comments}

Starting from a 5D Riemann-flat canonical metric on which we make
a dynamical foliation, in this letter we have studied a model for
a massless scalar field which drives a gravitational collapse of
the universe, which finally acquires a finite size:
$a_0=\frac{3}{2} \sqrt{{3\over \Lambda_0}}$ related to the
cosmological constant $\Lambda_0$, but with null asymptotic scalar
curvature. An important result here obtained is that the
asymptotic universe (absent of singularities) results to be finite
and absent of matter, with $\left.\omega\right|_{t\rightarrow
\infty} \rightarrow -\infty$. This is because the pressure is
negative (opposes the collapse) along all the contraction and its
asymptotic value tends to zero, but more slowly than does the
energy density. This seems puzzling but is the apparent result of
the continuous dispersal\cite{dis} of the universe during its
collapse. This topic deserves a further study. But the more
notorious result is that the asymptotic back-reaction effects are
given by a negative constant [see eq. (\ref{br})]. Finally, a
remarkable difference of this model with phantom scenarios is that
here the kinetic component of energy density related to the
massless scalar field remains always positive. The negative
contribution to the energy density, which in our model is the
responsible of the deceleration of the collapse, being given by
the back-reaction effects described in eq. (\ref{imp}), with an
asymptotical contribution given by (\ref{br}).

\section*{Acknowledgements}

\noindent MB acknowledges UNMdP and CONICET Argentina for
financial support.

\end{document}